# COLD ACCRETION DISKS WITH CORONAE AND ADVECTION

Xingming Chen

Department of Astronomy and Astrophysics, Göteborg University and Chalmers University of Technology, 412 96 Göteborg, Sweden; chen@fy.chalmers.se

## ABSTRACT

Cold optically thick accretion disks with hot coronae and radial advection have been investigated. Within the framework of $\alpha$-viscosity models, we assume that all the mass accretion and angular momentum transport take place in the cold disk, but that a fraction of the gravitational energy released is dissipated in the corona. Both the coronal energy dissipation and the advection heat transport have a stabilization effect on the thermal and viscous instabilities of the disk. If that more than $\sim 95$ percent of the total power is dissipated in the corona, then the locally unstable behavior of the disk is restricted to a relatively narrow spatial region and is found to lie in a small range of mass accretion rates. The global temporal variability of the disk can be very mild or may disappear, and which may be applicable to the low-frequency ($\sim 0.04$ Hz) quasi-periodic oscillations observed in black hole candidates Cyg X–1 and GRO J0422+32.

*Subject headings:* accretion, accretion disks — black hole physics — instabilities

## 1. INTRODUCTION

The spectral and temporal behavior of galactic black hole candidates (BHCs) and active galactic nuclei (AGNs) contains valuable information about the underlying physics of the accretion process. BHCs such as Cyg X-1 have very distinct characteristics. Cyg X-1 exhibits low and high states as defined by the X-ray flux in 1–10 keV band. The luminosity difference between these states is approximately an order of magnitude. In the high state, the spectrum consists of two components, namely, a relatively stable soft blackbody component and a highly variable hard power-law component. In the low state, the intensity shows rapid chaotic temporal fluctuations and the spectrum is characterized by a power-law component over an energy range from 1 keV to more than 100 keV. For a detailed review see Liang & Nolan (1984) and Tanaka & Lewin (1993). Characteristic spectral properties of AGNs include the so-called UV bump blackbody radiation, the power-law spectra of X-rays and the presence of Fe K$\alpha$ lines (see, e.g., Pounds et al 1990; Matsuoka et al 1990). It is now well accepted that the soft X-ray black body spectral component from BHCs and the UV black body radiation from AGNs are due to a cold optically thick accretion disk. On the other hand, the power-law spectra from these sources reflect the presence of optically thin hot matter.

There are two general scenarios proposed for the optically thin hot matter, namely, an entirely optically thin hot accretion disk or a hot corona plus an optically thick cold disk. Early work was published in mid-1970s, such as Shapiro, Lightman, & Eardley (1976) for disk models, and Liang & Price (1977) for corona models. The structure of coronae has been studied with different heating mechanisms for the coronal gas. For example, Compton scattering of X-rays from the boundary layer or the innermost regions of an accretion disk may lead to the formation of a hot corona above the disk and a wind ( Begelman, McKee, & Shield 1983). Recently, Haardt & Maraschi (1991) proposed a two-phase thermal disk-corona model for Seyfert galaxies. In this model, most of the thermal soft photons emitted from the optically thick cold disk escape from the system by passing through the optically thin hot corona. A small fraction of them are Compton upscattered in the corona by the high-temperature thermal electrons, and which produces the hard power-law spectral radiation. About half of this radiation is directed toward the disk (i.e., irradiation) and is reprocessed to emerge as blackbody radiation; while the remaining half escapes from the system and is observed. The corona is characterized by its Thomoson scattering optical depth, $\tau_{es}$, and the fraction of gravitational energy dissipated there, $\eta$. This energy is balanced by the cooling of inverse Compton scattering. In order to obtain a power-law spectrum with an index of $\sim 1$ observed from Seyfert galaxies, it is required that $\eta \sim 1$. Along this line, the structure of disk-corona system has been studied, for example, by Nakamura & Osaki (1993), Kusunose & Mineshige (1994) and Svensson & Zdziarski (1994) with different assumptions regarding such as the mass accretion and angular momentum transport taking place within the cold disk and corona, and the irradiation heating on the disk. For example, Svensson & Zdziarski (1994) assumed that mass accretion and angular momentum transport take place only in the disk, and no irradiation heating on the disk is included.

Optically thick accretion disks have been widely studied. It is well known that accretion disks may suffer thermal-viscous instabilities when radiation pressure is important. In a standard $\alpha$-model (Shakura & Sunyaev 1973), the viscous stress, $\tau$, is proportional to the total pressure, $p$, i.e., $\tau = -\alpha p$, where $\alpha$ is a constant. Disks are thermally and viscously unstable if the pressure is radiation dominated (Lightman & Eardley 1974; Shakura & Sunyaev 1976). The thermal-viscous instabilities may be stablized or reduced in strength, however, in various situations. For exmaple, if the viscous stress is proportional only to the gas pressure, $p_g$, i.e., $\tau = -\alpha p_g$, and $\alpha$ is a constant (see, e.g., Lightman & Eardley 1974; Coroniti 1981; Stella & Rosner 1984), then the disk is always stable. On the other hand, if $\alpha = \alpha_0 (H/R)^n$ (e.g., Meyer & Meyer-Hofmeister 1983), where $H$ is the local scale height of the disk, $R$ is the distance from the compact object, and $\alpha_0$ and $n$ are constants, then the disk may be mildly unstable (Milsom, Chen, & Taam 1994). The radial advection also has a stabilizing effect on the disk. This can be understood through the so-called S-shaped curve on the mass accretion rate and disk surface density plane, where the upper-branch represents the advection dominated solution. On the upper-branch, the disk is thermally and viscously stable. Detailed accretion disk models with advection have been constructed, for example, by Abramowicz et al (1988, 1995), Kato, Honma, & Matsumoto (1988), Chen & Taam (1993), Narayan & Popham (1993), Narayan & Yi (1994, 1995a,b) and Chen (1995). The stabilizing effect of coronal dissipation on the thermal-viscous instability of accretion disks has been studied by Ionson & Kuperus (1984) and Svensson & Zdziarski (1994).

In this paper, we study accretion disks with both radial advection and coronal dissipation included and investigate the instability behavior of such disks. In the next section we outline the assumptions and approximations underlying the models. The numerical results are presented in §3 and their possible relevance to the low-frequency ($\sim 0.04$ Hz) quasi-periodic oscillations (QPOs) observed in BHCs is discussed in the final section.

## 2. FORMULATION

We investigate a disk-corona system with radial advection included and focus on the structure and stability of the cold accretion disk instead of the hot corona above it. The system is assumed to be axisymmetric and non self-gravitating. The disk is optically thick and geometrically thin so that it can be described by the vertically integrated equations. The corona is hot and optically thin. Following Svensson & Zdziarski (1994), it is assumed that all the mass accretion and angular momentum transport take place in the cold disk. Thus in the steady state, the mass and angular momentum conservation equations can be written as

$$\dot{M} = -2\pi R \Sigma v_r, \quad (1)$$

and

$$\nu\Sigma = \frac{\dot{M}}{3\pi} f g^{-1}, \quad (2)$$

where $\dot{M}$, $\Sigma$, $v_r$, and $\nu$ are the mass accretion rate, the surface density, the radial velocity and the kinematic viscosity respectively. The correction terms are $f = 1 - 9\Omega_*/(\Omega r^2)$ and $g = -(2/3)(d\ln\Omega/d\ln R)$, where $r = R/R_G$, $\Omega_* = \Omega(3R_G)$, and $R_G = 2GM/c^2$ is the Schwarzschild radius. We assume a pseudo-Newtonian potential (Paczyński & Wiita 1980), $\Phi = -GM/(R - R_G)$, and the disk is rotating at the Keplerian velocity $\Omega = \sqrt{GM/R(R - R_G)^2}$. The viscosity is discribed with the standard $\alpha$-model prescription as (Shakura & Sunyaev 1973):

$$\nu = \frac{2}{3}\alpha c_s H, \quad (3)$$

where $\alpha$ is a constant and $c_s$ and $H$ are the local sound speed and the half-thickness of the disk respectively:

$$c_s = \sqrt{p/\rho}, \qquad H = c_s/\Omega_k. \quad (4)$$

Here $\rho = \Sigma/2H$ is the density and $p$ is total pressure. The steady state energy conservation equation for the cold accretion disk is represented by the balance between the local viscous heating, $Q_+$, the local radiative cooling in the vertical direction, $Q_-$, and the global heat transport (radial advection), $Q_{adv}$. It is expressed as

$$Q_+ = Q_- + Q_{adv}. \tag{5}$$

Here, irradiation heating from the corona has not been included (Svensson & Zdziarski 1994). This is justified by the fact that the heating of the disk by the corona only takes place in the surface layers down to a few Thompson optical depths (see Ross & Fabian 1993). We assume that a fraction $\eta$ of the total gravitational energy released is dissipated in the corona (Haardt & Maraschi 1991). Thus, within the disk, the viscous heating rate per unit area becomes

$$Q_+ = (1-\eta)\frac{3\dot{M}}{4\pi}\Omega^2 fg. \tag{6}$$

The local radiative cooling rate per unit area, in the optically thick radiative diffusion approximation, is

$$Q_- = \frac{4acT^4}{3\kappa\Sigma}, \tag{7}$$

where $T$ is the mid-plane temperature of the disk and $\kappa$ is the opacity assumed to be electron scattering, $\kappa_{es} = 0.34$. The advection cooling term is taken in a form (see, e.g. Chen & Taam 1993):

$$Q_{adv} = -\frac{\Sigma v_r}{R}\frac{p}{\rho}\xi = \frac{\dot{M}}{2\pi R^2}\frac{p}{\rho}\xi, \tag{8}$$

and

$$\xi = -\frac{4-3\beta}{\Gamma_3-1}\frac{d\ln T}{d\ln R} + (4-3\beta)\frac{d\ln \Sigma}{d\ln R}, \tag{9}$$

where $\beta = p_g/p$ is the ratio of the gas pressure to the total pressure, $\Gamma_3 = 1 + (4-3\beta)(\gamma-1)/[\beta + 12(\gamma-1)(\beta-1)]$, and $\gamma = 5/3$ is the ratio of specific heats. For optically thick accretion disks, the equation of state is given by

$$p = \frac{\mathcal{R}\rho T}{\mu} + \frac{aT^4}{3}, \tag{10}$$

where $\mu$ is the mean molecular weight assumed to be 0.617.

## 3. THERMAL EQUILIBRIA AND STABILITY

By replacing equation (9) with a constant value of $\xi$, the surface density and mid-plane temperature of the disk can be solved from a set of algebraic equations for a given mass accretion rate and a fixed radius. This is the procedure used by Abramowicz et al (1995) to calculate the thermal equilibrium curves of either optically thick (see also Abramowicz, Lasota, & Xu 1986) or optically thin accretion disks. In the former case, the S-shaped relation between the mass accretion rate and the surface density is close to that calculated from the full set of slim disk equations (see Abramowicz et al 1988; Chen & Taam 1993). In the later case, this procedure is also confirmed by

Chen (1995) to be very adequate. Accordingly, we use the same approach as that of Abramowicz et al (1986, 1995) to examine the $\dot{M}(\Sigma)$ relation. The numerical value of $\xi$ can be estimated when the advection cooling is a dominant process (only then it becomes important). In that case, by combining equations (2)–(4), (6) and (8), one has

$$\Sigma = 44.368 r^{-1/2} \dot{m} \alpha^{-1} g^{-2} q (1-\eta)^{-1} \xi, \tag{11}$$

where $q = 1 - r^{-1}$, $\dot{m} = \dot{M}/\dot{M}_c$, and $\dot{M}_c$ is the critical mass accretion rate defined as $\dot{M}_c = 64\pi GM/(c\kappa_{es})$. Here we have assumed 1/16 for the efficency of the conversion of rest-mass energy into the radiation. If we assume the disk is radiation pressure dominated then a relation between $T$ and $\Sigma$ can be derived from equations (2)–(4), and (10) as,

$$T^4 = 2.464 \times 10^{30} r^{-9/4} \Sigma^{1/2} \dot{m}^{1/2} \alpha^{-1/2} m^{-1} f^{1/2} g^{-1/2} q^{-3/2}, \tag{12}$$

where $m$ is the mass of the central object expressed in unit of the solar mass. If we let the correction terms, $f$, $g$, and $q$ be constants independent of radius, then $\Sigma \propto r^{-1/2}$ and $T \propto r^{-5/8}$. Thus, with $\beta = 0$, we have,

$$\xi = -12 \frac{d \ln T}{d \ln R} + 4 \frac{d \ln \Sigma}{d \ln R} = 5.5. \tag{13}$$

For definiteness we hereafter assume that $m = 10$, $\alpha = 0.1$, and $\xi = 5.5$.

We first present the thermal equilibria of accretion disks with different $\eta$, the fraction of energy dissipated in the corona (Fig. 1). For $\eta \lesssim 0.99$, there are three branches. The lower and middle branches represent the local radiative cooling solutions (see Svensson & Zdziarski 1994). The former one is gas pressure dominated and is thermally and viscously stable. The later one is radiation pressure dominated and is thermally and viscously unstable. The upper-branch represents the advection cooling solution, it is thermally and viscously stable (see Abramowicz et al 1988; Kato, Honma, & Matsumoto 1988; Chen & Taam 1993; Narayan & Popham 1993; Narayan & Yi 1994, 1995b) By increasing $\eta$, the efficiency of the cooling (both radiative and advection) is effectively increased by a factor of $(1-\eta)^{-1}$. The lower and upper branches move down-right since $Q_- < Q_+$ or $Q_{adv} < Q_+$ below these two lines. The middle-branch moves in the up-right direction since $Q_- < Q_+$ at the right side of this line. The net result is that, both the lower turning point and the upper turning point move to the up-right direction and $|\dot{M}_2 - \dot{M}_1|$ and $|\Sigma_1 - \Sigma_2|$ decrease. Here, $(\Sigma_1, \dot{M}_1)$ and $(\Sigma_2, \dot{M}_2)$ denote the surface densities and the mass accretion rates at these two turning points respectively. Furthermore, the middle-branch becomes steeper (more perpendicular to the $\Sigma$ axis) and eventually disappears for $\eta \gtrsim 0.99$. A steep middle-branch reduces the strength of the instability (see Milsom, Chen, & Taam 1994; Chen & Taam 1994). This result is mainly due to the advection, which can be illustrated through the dotted curves where no advection is included (Svensson & Zdziarski 1994). It is seen that the dotted lines do not turn to the right to form the upper-branches. The slopes of the dotted middle-branches are basically unchanged for various $\eta$ and they are flatter than the corresponding ones with advection included.

The global temporal variability of accretion disks depends not only on the local instability at a specified radius but also on the spatial size of the disk in which it is unstable (Chen & Taam

1994). For the later reason, the thermal equilibria of accretion disks at different radii are shown in Figure 2 for two cases with $\eta = 0$ and $\eta = 0.95$ respectively. For the former case, only radial advection but no coronal dissipation is included. It is seen that instability (the middle-branch) exists in wide ranges of both disk radii and mass accretion rates. This results in a global, burst-like time-evolution (see Taam & Lin 1984; Lasota & Pelat 1991). The inclusion of coronal dissipation apparently weakens the instability of the disk in three aspects (in comparision to $\eta = 0$). First, the spatial size where the disk is locally unstable is reduced to $r \lesssim 100$ from $r > 500$. Second, the range of mass accretion rates at which the disk is unstable is greatly reduced to a neighborhood of $\sim 0.1 - 1 \dot{M}_c$. Third, the strength of the local instability is also reduced (steeper slope). Thus the global temporal variability of the disk can be very mild or totally disappears for $\eta \gtrsim 0.95$.

## 4. DISCUSSION

We have demonstrated that accretion disks with coronae may have only a very mild thermal-viscous instability in comparision to models without a corona. This is due to the stabilization effects of coronal dissipation and radial heat transport within the cold disk. We suggest that this result is applicable to the low-frequency QPOs ($\sim 0.04$ Hz) observed in the low states of black hole candidates Cyg X-1 (Angelini & White 1992; Vikhlinin et al 1992a, 1994; Kouveliotou et al 1992a) and GRO J0422+32 (Vikhlinin et al 1992b, Kouveliotou et al 1992b, Pietsch et al 1993).

It is well-known that accretion disks which possess an S-Shaped thermal equilibrium curve in the mass accretion rate and surface density plane $(\Sigma, \dot{M})$ may experience a limit-cycle time evolution. This type of thermal-viscous instabilities has been widely studied by using nonlinear time-dependent calculations for accretion disks surrounding either white dwarfs or neutron stars and black holes. Even though the physics of the origin of these instabilities is different, their evolution behavior is quite similar (see Cannizzo 1993 for a review for dwarf novae and Taam & Lin 1994 and Lasota & Pelat 1991 for neutron stars). For example, for burst-like evolutions, it is understood that the recurrence time of bursts corresponds to the viscous diffusion time of the regions where the disk is unstable. On the other hand, the rising time of the burst corresponds to the thermal time scale and which is shorter than the viscous time scale. Thus, by reducing the radial width of the unstable disk the evolution may become oscillation-like once the thermal and viscous time scales become comparable. This is exactly the case as was shown by Chen & Taam (1994) based on a modified viscosity prescription. It was also shown there that the amplitude of the oscillation becomes smaller when the strength of the instability is weaker, which is represented by a steeper slope of the middle-branch of the thermal equilibrium curve. A type of small amplitude, periodic ($\sim 0.04$ Hz), oscillations has been obtained for both neutron star and black hole accretion disks and which was suggested to be relevant to the low-frequency QPOs observed (Chen & Taam 1994). For the present model, similar global time-dependent calculations are needed to constrain the disk-corona parameters, such as $\alpha$, $\eta$, $m$, and $\dot{M}$, quantitatively.

Nevertheless, the nature of the instability implies a time scale of $R^2/\nu \sim 10^{-5}\alpha^{-1}(R/H)^2 r^{3/2}$ s, which is in the order of tens of seconds for $r \sim 100$ and $H/R \sim 0.01 - 0.1$.

We have adopted the concept of coronal dissipation popularly applied in the recent two-phase disk-corona models (Haardt & Maraschi 1991). Ionson & Kuperus (1984) have used a slightly different approach. In their assumption, the disk consists of an internal and an external viscosities; only the internal one generates heating in the disk, while both of them transport the angular momentum. Since the physics of $\eta$ is not well understood, we leave it as a free parameter. We confirm the result of Ionson & Kuperus (1984) and Svensson & Zdziarski (1994) that coronal dissipation has a stabilizing effect on the thermal-viscous instability of accretion disks by raising the mass accretion rate threshold above which the disk is unstable.

We have not included the irradiation heating from the corona on the disk. The irradiation heating can dominate the viscous heating in the disk if it is deposited throughout the cold disk (see, e.g., Nakamura & Osaki 1993; Kusunose & Mineshige 1994). However, since the deposition and heating of the radiation from the corona only takes place in the surface layers down to a few Thompson depths (see, e.g., Ross & Fabian 1993), irradiation heating will probably not affect the cold optically thick disk very much (Svensson & Zdziarski 1994). Furthermore, it has been also known that the effect of irradiation is unimportant in the deep interior of the disk so long as the irradiated flux is much less than the product of the viscously generated flux and the optical depth of the disk (Lyutyi & Sunyaev 1976; Taam & Mészáros 1987). In the disk-corona system, the irradiation (from corona) and viscously generated (from disk) fluxes scale to $\eta/2$ and $1 - \eta$ respectively. Thus the criterion for unimportant irradiation reads $\kappa\Sigma \gg \eta/(1 - \eta)$. This is easily satisfied even for $\eta = 0.99$ due to a large surface density, $\Sigma \sim 10^5 - 10^6$.

The structure and stability of the corona is beyond the scope of the present paper. In fact, the corona should not be examined separately from the disk-corona system. In a self-consistent disk-corona model, one may need to consider the disk-corona system together and should probably address problems such as the radiation transfer and its two-dimension (radial-vertical) nature. Futhermore, the non-Keplerian effect and the transonic nature of accretion flows near the vicinity of the black hole will also play important roles in determining the global structure of the disk. These will be the subjects of our future investigations.

The author would like to thank Profs. Marek Abramowicz and Ronald Taam for discussions and also an anonymous referee for useful suggestions.

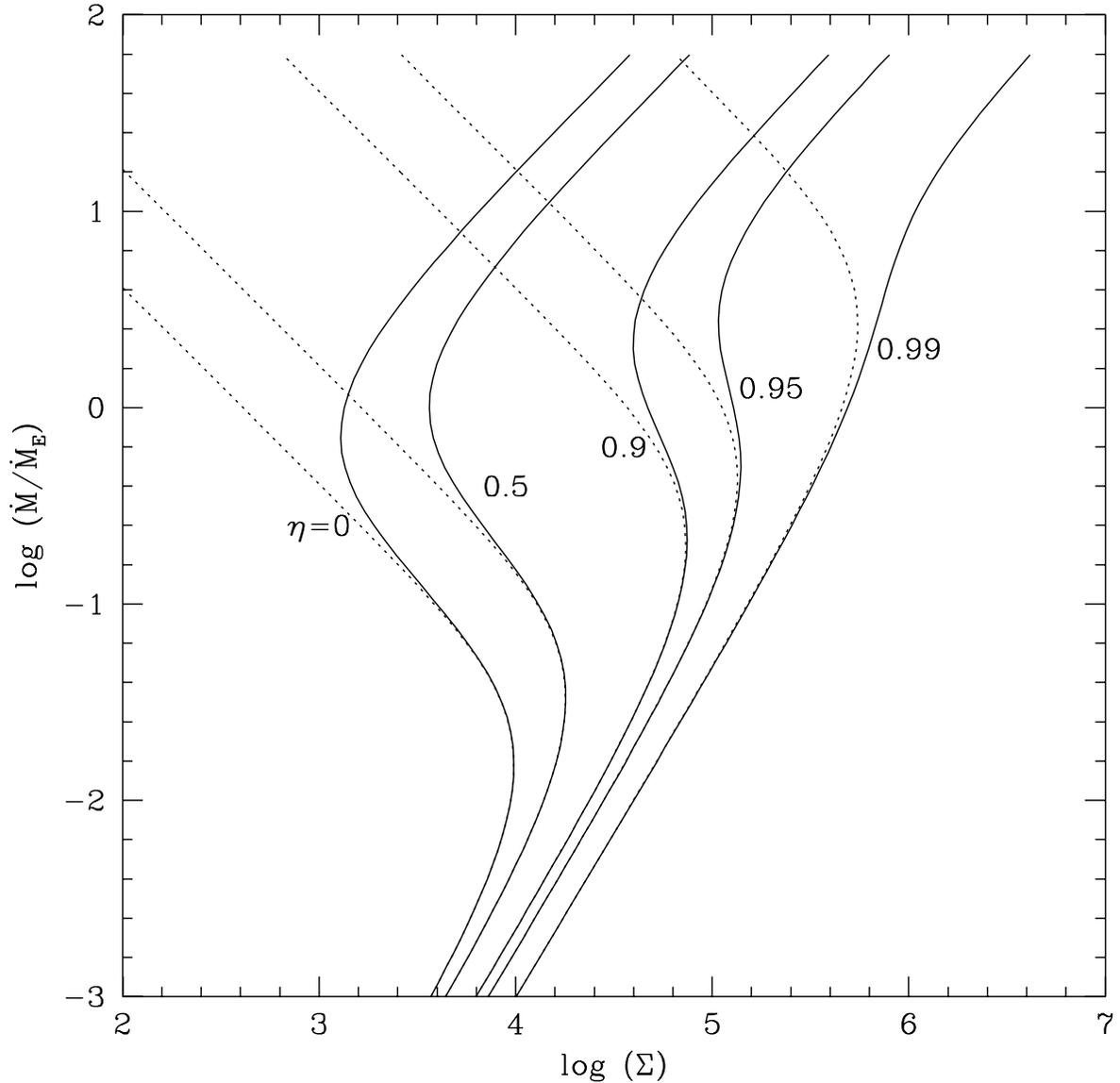

Fig. 1.— The S-shaped curves (solid lines) of $\dot{M}(\Sigma)$ relation for various $\eta$, the fraction of energy dissipated in the corona. Here $r = 10$, $\alpha = 0.1$, and $m = 10$. Note that for $\eta \gtrsim 0.99$, the entire disk is locally thermally and viscously stable. The dotted lines represent the solutions without advection.

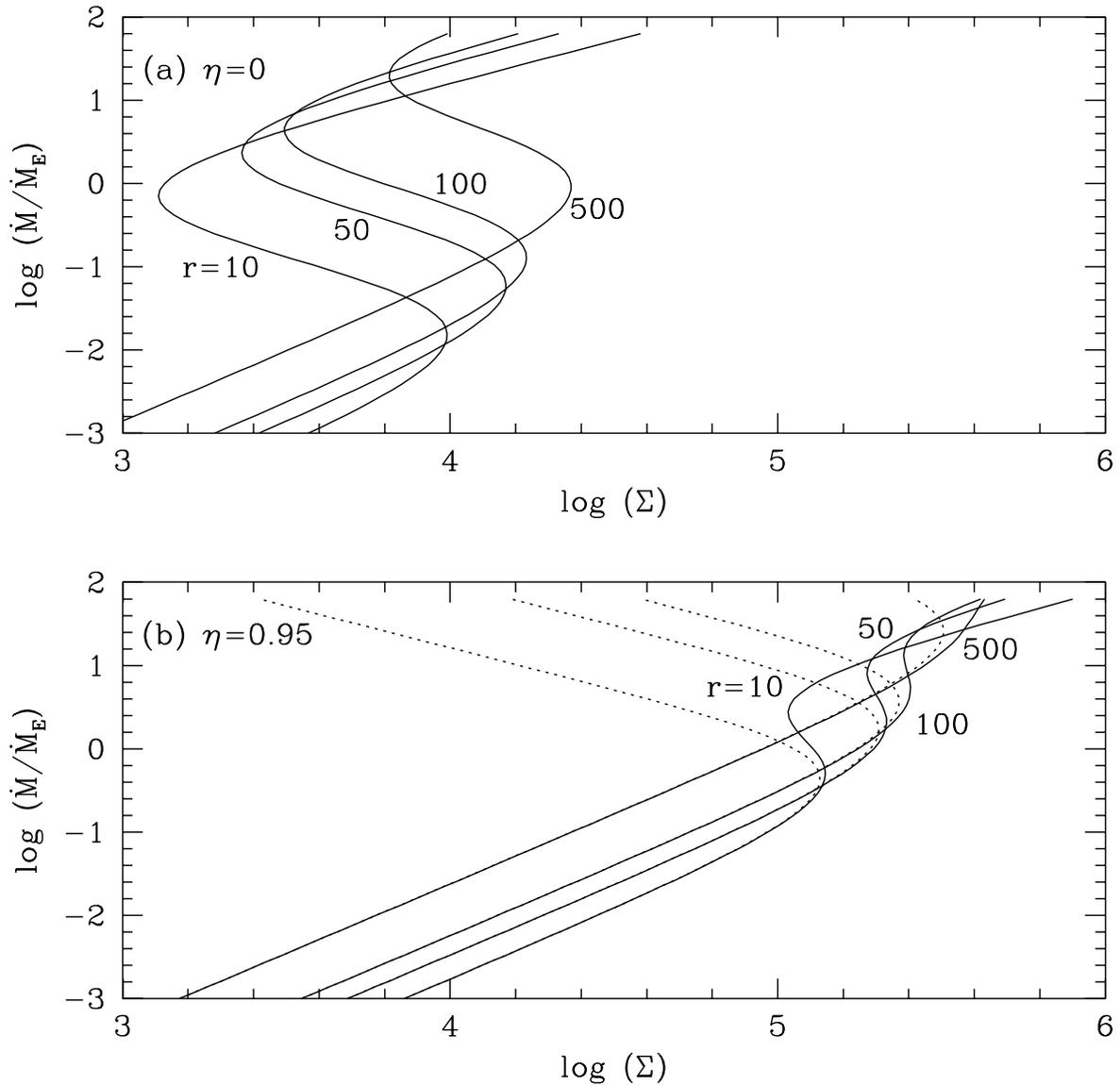

Fig. 2.— The S-shaped curves of $\dot{M}(\Sigma)$ relation at different radii. Note that for $\eta = 0.95$, the unstable behavior of the disk is restricted to a relatively narrow spatial region and is present in a small range of mass accretion rates. The dotted lines in b) represent the solutions without advection.